\documentclass[aps,pra,twocolumn,superscriptaddress]{revtex4-1}
\usepackage{graphicx}
\usepackage{epstopdf}
\usepackage{color}
\usepackage{bm}
\usepackage{amsmath}
\usepackage{float}
\usepackage[normalem]{ulem}
\begin{document}

\title{Mesonic excited states of magnetic monopoles in quantum spin ice}

\author{Olga Petrova}
\affiliation{D\'{e}partement de Physique, Ecole Normale Sup\'{e}rieure / PSL Research University, CNRS, 24 rue Lhomond, 75005 Paris, France}
\affiliation{Max Planck Institute for the Physics of Complex Systems, 01187 Dresden, Germany}

\begin{abstract}

Spin ice magnetic monopoles are fractionalized emergent excitations in a class of frustrated magnets called spin ices. The classical spin ice model has an extensive number of ground state spin configurations, whereas magnetic monopoles can be thought of as the endpoints of string operators applied to these ground states. Introducing quantum fluctuations into the model induces monopoles with dynamics, which would normally  lift the degeneracy of the two-monopole energy level at the linear order of the perturbation. Contrary to this expectation, we find that quantum fluctuations in the form of a locally transverse field term partially preserve the extensive degeneracy of the monopole pairs up to the much weaker splitting of the background monopole-free spin ice configurations. Each of these approximately degenerate excited states, termed \emph{mesons}, can be represented as a bound monopole-antimonopole pair delocalized in a classical spin ice background.
\end{abstract}

\maketitle

\section{Introduction}

Spin ice has attracted the attention of physicists from its early days \cite{Bramwell16112001}. One of the particularly exciting finds was the presence of gapped fractionalized excitations that behave as emergent magnetic monopoles \cite{Castelnovo}. In the recent years, the field has been revitalized by the ever growing quantum spin liquid community. Various aspects of introducing quantum fluctuations into the classical spin ice model have been studied, focusing on two main avenues: the emergence of the gapless quantum spin liquid ground state \cite{PhysRevB.69.064404,Henley,PhysRevB.86.104412, PhysRevB.86.075154,PhysRevB.92.094418,PhysRevB.95.134439,PhysRevB.96.125145} and the properties of magnetic monopole excitations 
\cite{PhysRevLett.108.247210,PhysRevB.92.100401,PhysRevLett.116.167202,PhysRevB.94.104401,PhysRevB.95.041106,PhysRevB.96.035136}. The latter route remains less explored, despite its broad relevance to some of the community's most sought after questions, such as how to identify a quantum spin liquid in an experiment \cite{PhysRevLett.98.157204,PhysRevLett.98.157204,PhysRevLett.99.237202,PhysRevLett.105.077203,PhysRevLett.109.017201,PhysRevLett.106.187202,PhysRevX.1.021002,NatCommun54970,Nakatsuji,
Sibille2018,doi:10.7566/JPSJ.87.064702} and what happens to a doped quantum dimer model \cite{Anderson1987}.

The classical picture of the magnetic monopoles arising in spin ice is rather straightforward. Monopoles come in pairs of ``magnetic charges" of the opposite sign -- sources and sinks of the magnetization. Each pair is associated with a finite energy cost, in addition to the effective Coulomb interaction between the monopoles. Introducing quantum fluctuations into the model induces dynamics of the magnetic monopoles, splitting their classical energies (highly degenerate due to the extensive number of possible spin ice backgrounds) into continuous bands. In the perturbative treatment of the quantum problem, this splitting happens at the linear order in the fluctuations, in contrast to the ground state spin ice manifold, whose degeneracy is only lifted by the higher order ring exchange terms.

In this article, we study the spin ice model with a local transverse field. Contrary to the aforementioned considerations, we find that quantum fluctuations of this form partially preserve the extensive degeneracy of the first excited energy level up to the order at which the ground state manifold splits. This results in an approximate flat band in the energy spectrum. This remarkable finding is a consequence of the peculiar structure of the model's state graph. The corresponding flat band eigenstates, termed \emph{mesonic} states, resemble a state graph counterpart to the Aharonov-Bohm cages \cite{PhysRevLett.81.5888} supported by certain tiling geometries. Interestingly, unlike Aharonov-Bohm cages, the spin ice mesons are completely delocalized in real space.

In the balance of the paper we present the construction of an excited mesonic state starting from an arbitrary classical spin ice configuration. The resulting wavefunction is an exact eigenstate of the zero and two monopole sectors of our quantum spin ice model, whether the latter is defined on a three-dimensional pyrochlore or a planar square lattice. Additionally, we discuss a peculiar property of the planar spin ice mesons: given that the splitting of the spin ice ground state manifold is neglected, the dynamic structure factor of the mesons is given by the classical spin ice correlator, with momentum shifted to the center of the next Brillouin zone.

One of the major topical questions facing condensed matter physics is how to identify a quantum spin liquid compound when we see one. Exhibiting no conventional long-range order, these phases present a special challenge to experimentalists. The unique nature of their excitations may offer a more fruitful approach than probing the featureless ground state. Normally introducing quantum fluctuations into a classical Hamiltonian washes out the sharp features, which is why the approximately flat meson band that we find for the quantum spin ice with a transverse field term is of particular interest. We determine that in a neutron scattering experiment, the fraction of the total scattered weight that is saturated by the mesons vanishes. However, the question of whether mesons can give rise to a sharp experimental signature of a different nature in an appropriate compound, remains open.

\section{The model}

The model we consider is that of nearest neighbor spin ice, with the addition of a locally transverse field $\bf t$:
\begin{equation}
H_\mathrm{spin}=\sum_{\langle ij\rangle}J S^{z}_i S^{z}_j+\sum_{i}t\cdot S^{x}_i,
\label{eq:SI}
\end{equation}
where $\vec{S}_i$ are spins $1/2$ living on the sites of the pyrochlore lattice. Their $z$ components are aligned along the local easy axis joining centers of neighboring tetrahedra \cite[and references therein]{Bramwell16112001,spinicereview}. The centers of the tetrahedra form a diamond lattice. The convention for the spin variables ${\bf S}_i$ is such that $S^{z}_{i}=+1$ when the spin is pointing from sublattice A to sublattice B of the diamond lattice, and $S^{z}_{i}=-1$ otherwise. The transverse field $\bf t$, oriented perpendicular to the local easy axis, introduces quantum fluctuations in the form of single spin flips. 

When $t=0$, any spin configuration with two spins pointing into, and two spins pointing out of, the center of each tetrahedron is a ground state of the Hamiltonian (\ref{eq:SI}). The two-in two-out condition is known as the \emph{ice rule}. Flipping a spin in a ground state results in the ice rule being violated at two adjacent tetrahedra. Flipping a string of spins pointing head-to-tail fractionalizes this excitation into two point-like quasiparticles, located at the string's endpoints, that behave as monopoles endowed with magnetic charge of opposite sign. New \emph{charge} degrees of freedom, living on the diamond lattice sites, are defined in the following manner: each spin pointing in (out) of a tetrahedron brings in a $+\frac{1}{2}$ ($-\frac{1}{2}$) unit of magnetic charge to the respective diamond lattice site. The ice rule for the spin variables translates to the zero net charge condition for the diamond lattice sites. The 3-in 1-out and 3-out 1-in ice rule violations then correspond to $+1$ and $-1$ magnetic monopoles respectively. Creation of higher energy $\pm2$ monopoles is energetically suppressed, and we will be excluding such configurations from our analysis. In particular, this restriction means that for a given magnetic monopole in spin ice, there are three, rather than four, ways for it to hop onto the nearest neighbor tetrahedron, following a spin flip.

Quantum fluctuations induce dynamics of the magnetic monopoles. Considering the model given in Eq. (\ref{eq:SI}), monopoles start to hop to adjacent tetrahedra at linear order in $t$, whereas the degeneracy of the monopole-free ground states is lifted at $\mathcal{O}(t^6/J^5)$. The latter process involves creation of a monopole pair at a cost proportional to $J$, then having the monopoles traverse a closed hexagonal loop, and get annihilated at the end. Therefore, quantum fluctuations have a parametrically larger effect on the magnetic monopole excitations than they do on the spin ice ground states. In this article, we choose to focus on the former effect, and neglect the splitting of the ground state manifold.

\section{Mesonic eigenstates}

\subsection{Construction of the mesonic wavefunctions}

\begin{figure}
\includegraphics[width=\linewidth]{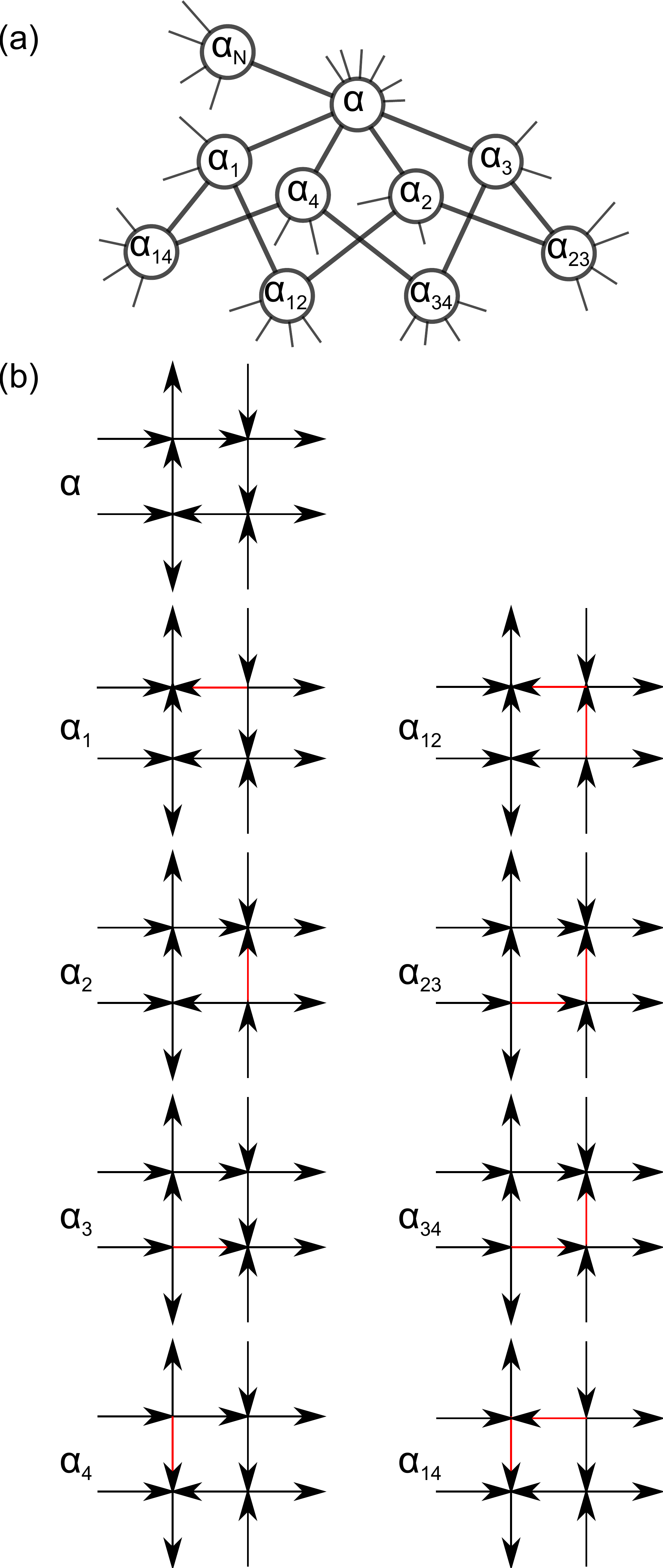}
\caption{(a) Fragment of the state graph of the model whose Hamiltonian is given by Eq.~(\ref{eq:SI}). Graph nodes labeled as $\alpha_i$ correspond to basis states $|\alpha_{\vec{r}_i}\rangle$. (b) Planar spin ice configurations corresponding to nodes of the state graph.}
\label{fig:stategraph}
\end{figure}

Consider the outcome of acting with the Hamiltonian (\ref{eq:SI}) on an arbitrary classical spin configuration labeled $\alpha$. In a system containing $N$ spins, the transverse field term is going to connect the basis state $|\alpha\rangle$ to $N$ states $|\alpha_{\vec{r}}\rangle$ 
\begin{equation}
|\alpha_{\vec{r}}\rangle=S^{x}_{\vec{r}}|\alpha\rangle,
\end{equation}
each containing a pair of magnetic monopoles located at the tetrahedra sharing the pyrochlore lattice site positioned at $\vec{r}$. If we restrict ourselves to the zero and two monopole sectors of Hilbert space, further applications of Hamiltonian (\ref{eq:SI}) will either connect states $|\alpha_{\vec{r}}\rangle$ back to monopole free spin ice $|\alpha\rangle$, or make one of the monopoles hop in one of two possible directions. Two-monopole states, where the monopoles are further than one diamond lattice spacing away from each other, are each connected to six other states, since each of the two monopoles has three ways to go. Graphically the structure of the Hamiltonian (\ref{eq:SI}) can be depicted via a so-called \emph{state graph}. Nodes of the state graph, shown in Fig.~\ref{fig:stategraph}(a), correspond to the basis states, i.e. classical spin configurations, depicted in Fig.~\ref{fig:stategraph}(b). Edges connect the nodes, whose corresponding states are connected by the Hamiltonian (\ref{eq:SI}). Unlike in the case of a diluted spin ice system \cite{PhysRevB.92.100401}, where one of the monopoles is fixed at the vacancy defect and the shortest closed loop in the state graph has length 20, the state graph of two propagating monopoles has numerous closed loops of length 4. These closed loops do not correspond to closed cycles on the physical lattice. In fact, monopole's going around a closed cycle (the shortest being a hexagon or a square on diamond and square lattices respectively) connects two distinct spin ice states $|\alpha\rangle$ and $|\beta\rangle$.

\begin{figure}
\includegraphics[width=0.95\linewidth]{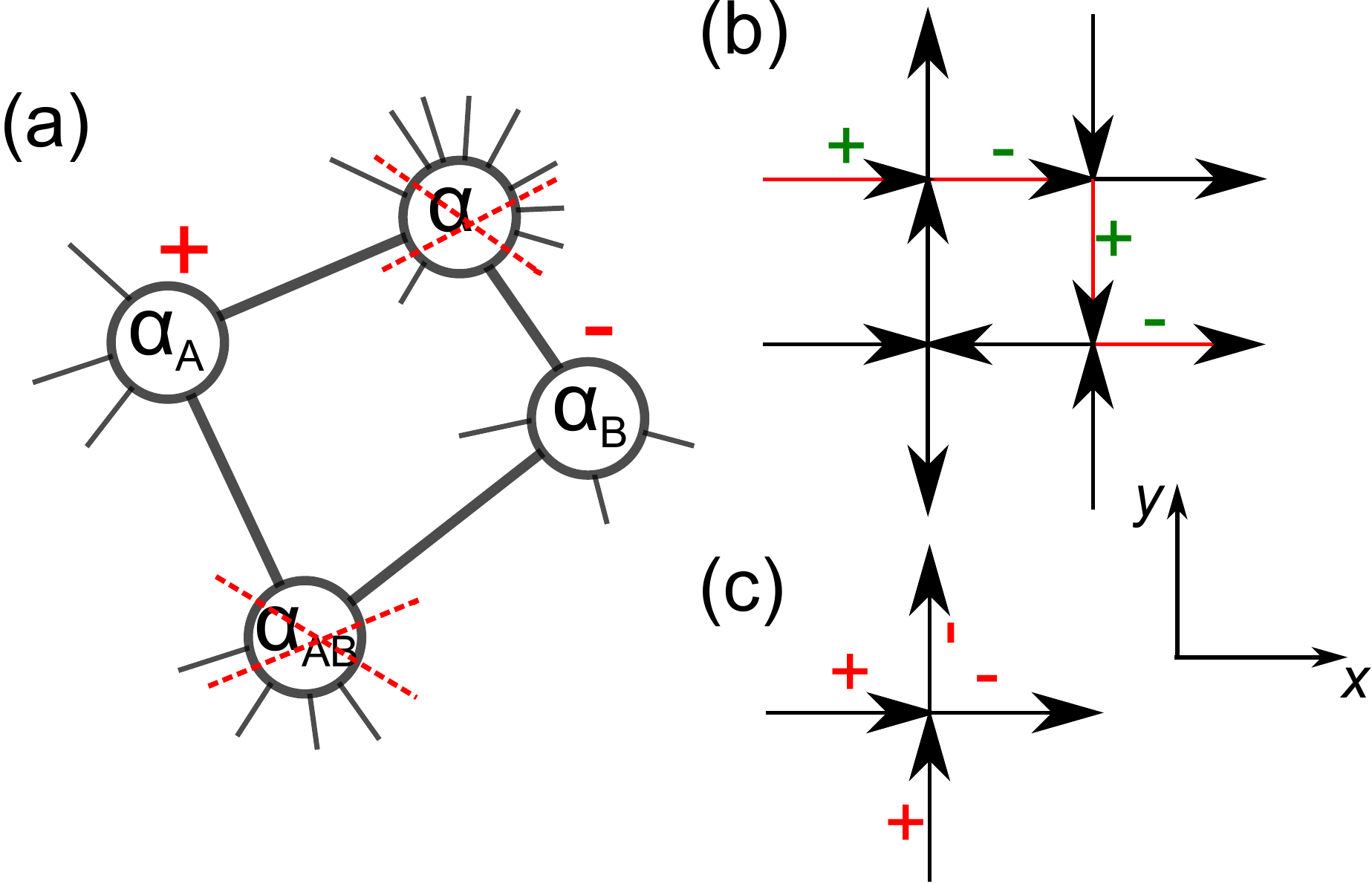}
\caption{(a) Antisymmetric combination of $|\alpha_{A}\rangle$ and $|\alpha_{B}\rangle$ cancels out the hopping to both $|\alpha_{AB}\rangle$ and $|\alpha\rangle$. (b) Mesonic phase factors for basis states $|\alpha_{\vec{r}}\rangle$, indicated in green for a given spin at $\vec{r}$, are assigned in an alternating manner along a head-to-tail string of spins in $|\alpha\rangle$, shown in red. (c) There are two head-to-tail strings going in and out of every junction, allowing for consistent assignment of the meson phase factors.}
\label{fig:shell}
\end{figure}

Let us consider the nodes of the first and second shells of the state graph, centered around an arbitrary monopole-free node $\alpha$. Nodes of the first shell are labeled by the position of a spin, flipped relative to the reference state  $|\alpha\rangle$. Each node of the second shell has two labels, corresponding to two flipped adjacent spins, that are pointing head-to-tail. Since there is a choice of which of the two spins is flipped first, each of the second shell nodes is connected to exactly two first shell nodes. This is shown in Fig.~\ref{fig:shell}(a) where the second shell node $\alpha_{AB}$ is connected to $\alpha_{A}$, $\alpha_{B}$, and four more nodes on the third shell, that correspond to the monopoles being taken further apart from one another. We can now attempt to construct an eigenstate of the Hamiltonian (\ref{eq:SI}), in which only spin configurations corresponding to the nodes on the first shell have non-zero amplitudes. For this, it is convenient to think of the model as a single particle hopping on the state graph with hopping amplitude $t$. All nodes except for those corresponding to monopole-free spin ice configurations have potential equal to the energy cost of a single monopole pair. If we assign equal amplitudes, and opposite $\pm$ phase factors to nodes (i.e., spin configurations)  $\alpha_{A}$ and $\alpha_{B}$, hopping to both  $\alpha_{AB}$ and  $\alpha$ is canceled out. However, both $\alpha_{A}$ and $\alpha_{B}$ are each connected to three more nodes in the second shell, which, in turn, are connected to other first shell nodes. Indeed, in order to cancel hopping to all of the second shell nodes, we need to assign amplitudes with appropriate $\pm$ signs to all of the nodes of the first shell. The recipe for assigning these phase factors turns out to be simple: we choose an arbitrary one-dimensional path, going through all spins in the system along a head-to-tail direction, as shown in red in Fig.~\ref{fig:shell}(b), in a classical monopole-free spin ice configuration $\alpha$. In order to construct a quantum eigenstate, we assign equal amplitudes, and alternating $\pm$ phase factors to the spin configurations corresponding to all the nodes $\alpha_{\vec{r}}$ as we follow the path, where $\vec{r}$ is the position of a spin, belonging to the path, on the physical lattice. We can check that such sign assignments can always be done in a consistent manner by considering what happens at a given tetrahedron (or vertex of the square lattice for planar spin ice) [Fig.~\ref{fig:shell}(c)], where there are two possible path directions to consider.

The resulting eigenstate, termed \emph{mesonic}, is a superposition of all single spin flips in a given monopole-free spin ice background. The meson is an exact eigenstate of the Hamiltonian (\ref{eq:SI}) in the zero and two monopole sectors, with energy equal to the cost of a single monopole pair introduced into a spin ice system. Remarkably, we can construct as many mutually orthogonal mesons, as there are classical spin ice configurations. If we treat the transverse field term in the Hamiltonian (\ref{eq:SI}) as a perturbation, the Pauling degeneracy of the mesonic energy level is lifted at the same order as the degeneracy of the ground state manifold.

Let us write down the mesonic wavefunction for planar spin ice, using the method that we established above for assigning the phase factors. For a given classical spin ice configuration $|\alpha\rangle$, the corresponding normalized mesonic state $|m_{\alpha}\rangle$ is given by:
\begin{equation}
|m_{\alpha}\rangle=\sum_{{\vec r}}\cfrac{(-1)^{r_x+r_y}}{\sqrt{N}}\sigma^{x}_{\vec r}\sigma^{z}_{\vec r}|\alpha\rangle,
\label{eq:meson2D}
\end{equation}
where ${\vec r}$ are positions of the spins, and $\vec{\sigma}$ are the global spin variables. The convention for the latter is that $\sigma^{z}=+1$ when the horizontal/vertical spin is pointing in the direction of the $x/y$ axis, oriented as shown in Fig.~\ref{fig:shell}. The expression Eq.~(\ref{eq:meson2D}) can be rewritten in terms of local spin variables $\vec{S}$ that were used in Eq. (\ref{eq:SI}):
\begin{equation}
|m_{\alpha}\rangle=\cfrac{1}{\sqrt{N}}\sum_{{\vec r}}S^{x}_{\vec r}S^{z}_{\vec r}|\alpha\rangle.
\label{eq:localmeson}
\end{equation}
This form of the mesonic wavefunction  (\ref{eq:localmeson}) also applies to the three-dimensional pyrochlore spin ice model. Let us verify that these are indeed exact eigenstates of the Hamiltonian (\ref{eq:SI}) in the zero and two monopole sectors. In order to do this, we show that mesons (\ref{eq:localmeson}) have zero kinetic energy as long as creation of additional monopole pairs is suppressed:
\begin{gather}
    t\sum_{\vec{R}}S^{x}_{\vec{R}}|m_{\alpha}\rangle=\cfrac{t}{\sqrt{N}}\sum_{\vec{R},\vec{r}}S^{x}_{\vec{R}}S^{x}_{\vec r}S^{z}_{\vec r}|\alpha\rangle= \nonumber \\ 
   \cfrac{t}{\sqrt{N}}\left(\sum_{\vec{r}}S^{z}_{\vec r}|\alpha\rangle+\sum_{\vec{r}>\vec{R}}S^{x}_{\vec{R}}S^{x}_{\vec r}(S^{z}_{\vec R}+S^{z}_{\vec r})|\alpha\rangle\right)=0, \nonumber
\end{gather}
where the restriction on the number of monopoles means that $\vec{R}$ and $\vec{r}$ are either equal to one another, or are adjacent and pointing head-to-tail (in which case, $S^{z}_{\vec R}|\alpha\rangle=-S^{z}_{\vec r}|\alpha\rangle$).

\subsection{Dynamic structure factor}

The dynamic structure factor is a quantity that can be measured directly in inelastic neutron scattering experiments, thus serving as a valuable probe of the spin correlations. For neutrons polarized in the $x$ direction, it is given by
\begin{equation}
S^{xx}(\vec{q},\omega)=\sum_{fi}\delta(E_f-E_i-\omega)\left|\sum_{\vec{R}}\langle f|\sigma^{x}_{\vec{R}}|i\rangle e^{i\vec{q}\cdot\vec{R}} \right|^2,
\label{eq:strucfac}
\end{equation}
where $\vec{q}$ and $\omega$ are momentum and energy transferred from the incoming neutron, $i$ can be thought of as the initial state(s) of the system, and sum over $f$ is a sum over all states. Because we choose to focus on the most visible effects of the quantum fluctuations, we neglect the splitting of the spin ice ground state manifold. In this approximation, to the leading order in the transverse field $t$ the sum over the actual ground states $|i\rangle$, which are quantum eigenstates of the Hamiltonian (\ref{eq:SI}), can be approximated by the sum over the classical spin ice configurations $\alpha$. Only the meson $|m_\alpha\rangle$ that is constructed from a given ice configuration $\alpha$ will have a non-zero overlap with the state corresponding to a single spin flip in the said configuration. Therefore, when estimating the meson contribution $S_{M}^{xx}(\vec{q})$ to the dynamic structure factor at $\omega$ equal to the energy cost of a single spin flip, the double sum over the quantum states $i$ and $f$ in Eq. (\ref{eq:strucfac}) becomes the sum over the classical spin ice configurations:
\begin{equation}
S_{M}^{xx}(\vec{q})=\sum_{\alpha}\left|\sum_{\vec{R}}\langle m_\alpha|\sigma^{x}_{\vec{R}}|\alpha\rangle e^{i\vec{q}\cdot\vec{R}} \right|^2.
\label{eq:strucfac1}
\end{equation}
We proceed to calculate mesonic structure factor for the planar spin ice using Eq. (\ref{eq:meson2D}):
\begin{widetext}
\begin{align}
S_{M}^{xx}(\vec{q})& = &\cfrac{1}{N}\sum_{\alpha}\left|\sum_{\vec{R}}\sum_{\vec{r}}\langle\alpha|(-1)^{r_x+r_y}\sigma^{z}_{\vec{r}}\sigma^{x}_{\vec{r}}\sigma^{x}_{\vec{R}}|\alpha\rangle e^{i\vec{q}\cdot\vec{R}} \right|^2 & = & \cfrac{1}{N}\sum_{\alpha}\left|\sum_{\vec{R}}\sum_{\vec{r}}\delta_{\vec{r},\vec{R}}\langle \alpha|(-1)^{r_x+r_y}\sigma^{z}_{\vec{r}}|\alpha\rangle e^{i\vec{q}\cdot\vec{R}} \right|^2 &=\nonumber \\ \nonumber
&& \cfrac{1}{N}\sum_{\alpha}\left|\sum_{\vec{R}}\langle \alpha|(-1)^{R_x+R_y}\sigma^{z}_{\vec{R}}|\alpha\rangle e^{i\vec{q}\cdot\vec{R}} \right|^2 & = &\cfrac{1}{N} \sum_{\alpha}\left|\sum_{\vec{R}}\langle\alpha|\sigma^{z}_{\vec{R}}|\alpha\rangle e^{i(\vec{q}+(\pi,\pi))\cdot\vec{R}} \right|^2 & = \nonumber \\ 
&&\cfrac{1}{N}\sum_{\alpha}\sum_{\vec{R}}\sum_{\vec{r}}\langle\alpha|\sigma^{z}_{\vec{R}}\sigma^{z}_{\vec{r}}|\alpha\rangle e^{i(\vec{q}+(\pi,\pi))\cdot(\vec{R}-\vec{r})}&=&
\sum_{\alpha}\sum_{\vec{R}}\langle\alpha|\sigma^{z}_{\vec{0}}\sigma^{z}_{\vec{R}}|\alpha\rangle e^{i(\vec{q}+(\pi,\pi))\cdot\vec{R}}.
\label{eq:planarstruc}
\end{align}
\end{widetext}
Remarkably, the last part of Eq. (\ref{eq:planarstruc}) is nothing but the static structure factor $S^{zz}(\vec{q})$ of the classical spin ice, with momentum $\vec{q}$ shifted by $(\pi,\pi)$. 

In order to investigate whether mesons can give rise to an appreciable signature in a neutron scattering experiment, we calculate the fraction of the scattering intensity that is saturated by the mesons. We estimate the total scattering intensity by considering spin ice in Eq. (\ref{eq:SI}) in the absence of quantum fluctuations. The sum over the initial states $i$ is then again taken over the classical spin ice configurations $\alpha$. For each $\alpha$, there are $N$, and only $N$, excited states $|f^{\alpha}_{\vec r}\rangle=\sigma^{x}_{r}|\alpha\rangle$, such that $\langle f|\sigma^{x}_{r}|\alpha\rangle\neq0$. The dynamic structure factor Eq. (\ref{eq:strucfac}) then takes the form
\begin{align}
S^{xx}_{t=0}(\vec{q})&=\sum_{i}\sum_{f}\sum_{{\vec R},{\vec R}'} e^{i{\vec q}({\vec R}-{\vec R}')} \langle i|\sigma^{x}_{\vec{R}'}|f\rangle \langle f|\sigma^{x}_{\vec{R}}|i\rangle= \nonumber \\
&\sum_{\alpha}\sum_{r}\sum_{{\vec R},{\vec R}'} e^{i{\vec q}({\vec R}-{\vec R}')} \langle \alpha|\sigma^{x}_{\vec{R}'}\sigma^{x}_{\vec{r}}|\alpha\rangle \langle \alpha|\sigma^{x}_{\vec{r}}\sigma^{x}_{\vec{R}}|\alpha\rangle = \nonumber \\
&\sum_{\alpha}\sum_{r}\sum_{{\vec R},{\vec R}'} e^{i{\vec q}({\vec R}-{\vec R}')} \delta_{{\vec R}',{\vec r}}\delta_{{\vec R},{\vec r}} = \sum_{\alpha} N.
\label{eq:strucfacCSI}
\end{align}

To estimate the fraction of the scattering intensity that is saturated by the mesons, we sum $S_{M}^{xx}(\vec{q})$ and $S^{xx}_{t=0}(\vec{q})$ over the N momenta ${\vec q}$, and find their ratio:
\begin{align}
\cfrac{\sum_{\vec{q}}\sum_{\alpha}\sum_{\vec{R}}\langle\alpha|\sigma^{z}_{\vec{0}}\sigma^{z}_{\vec{R}}|\alpha\rangle e^{i(\vec{q}+(\pi,\pi))\cdot\vec{R}}}{\sum_{\vec{q}}\sum_{\alpha} N}= \nonumber \\
\cfrac{\sum_{\alpha}\sum_{\vec{R}}\langle\alpha|\sigma^{z}_{\vec{0}}\sigma^{z}_{\vec{R}}|\alpha\rangle\delta_{{\vec R},0}}{\sum_{\vec{q}}\sum_{\alpha} N}=\cfrac{\sum_{\alpha}N}{\sum_{\vec{q}}\sum_{\alpha}N}=\cfrac{1}{N}.
\end{align}
Since the scattering weight that is saturated by the mesons scales as the inverse of the system's volume, we conclude that there is little hope of detecting them in a neutron scattering experiment. However, it is possible that the extensively degenerate flat band may be observed in other ways, particularly in a regime where the amplitude of the quantum fluctuations is small enough to neglect the ring exchange terms splitting the ground state manifold, yet strong enough to give sufficient bandwidth to the dispersive states.

\section{Conclusion}

In this article, we construct a family of nearly exact excited states for exchange quantum spin ice in a locally transverse field. These states, called mesons, have a number of peculiar properties. One meson can be constructed for each classical spin ice configuration. Up to the splitting of the ground state manifold, all mesons are degenerate with the energy equal to that of a single spin flip in the classical spin ice. The mesonic wavefunctions only have non-zero amplitudes at the single-spin flip basis states. Although we determine that in a neutron scattering experiment, the fraction of the total scattered weight that is saturated by the mesons vanishes, mesons may be visible in a different experimental setup.

The author thanks Roderich Moessner and S. L. Sondhi for their invaluable guidance during my time at MPI PKS, and Frank Pollmann, Siddhardh C. Morampudi, Sreejith Ganesh Jaya, Wladimir Tschischik, Oleg Tchernyshyov, Yuan Wan, Benoit Doucot, Nicolas Regnault, and Maurizio Fagotti for useful discussions. This work was supported by LabEX ENS-ICFP: ANR-10-LABX-0010/ANR-10-IDEX-0001-02 PSL*.

\bibliography{spinicebib}

\end{document}